\documentclass[aps,prd,preprint,longbibliography]{revtex4-1}

\usepackage{units}
\usepackage{amsmath,amssymb,amsthm,amstext,amsfonts}
\usepackage{graphicx}
\usepackage[pdftex,plainpages=false,pdfpagelabels]{hyperref}

\begin{document}

%\linenumbers

\noindent Phys. Rev. D 96, 116011 (2017)
\hfill arXiv:1710.02507 %%\,(\version)
%
%arXiv:1710.02507 \hfill KA--TP--31--2017\,(\version)
\vspace*{4mm}
%\preprint{KA--TP--31--2017\,(\version)}\vspace*{2mm}

\title{Improved bound on isotropic Lorentz violation
in the photon sector from extensive air showers}

\author{F.R. Klinkhamer}
\email{frans.klinkhamer@kit.edu}
\affiliation{
\mbox{Institute for Theoretical Physics,
Karlsruhe Institute of Technology (KIT),}\\
76128 Karlsruhe, Germany}
\author{M. Niechciol}
\email{niechciol@physik.uni-siegen.de}
\author{M. Risse}
\email{risse@hep.physik.uni-siegen.de}
\affiliation{
\mbox{Department of Physics, University of Siegen,}\\
57068 Siegen, Germany\vspace*{2mm}}

\begin{abstract}
\vspace*{2mm}
Cosmic rays have extremely high particle energies
(up to $10^{20} \; \text{eV}$) and can be used to search for
violations of Lorentz invariance. We consider isotropic
nonbirefringent Lorentz violation in the photon sector for
the case of a photon velocity larger than the maximum
attainable velocity of the standard fermions. Up to now,
Earth-based bounds on
this type of Lorentz violation have been determined from
observations of TeV gamma rays.
Here, we elaborate on a novel approach to test Lorentz
invariance with greatly improved sensitivity. This approach is based on
investigating extensive air showers which are induced by cosmic-ray
particles in the Earth's atmosphere. We study the impact of
two Lorentz-violating decay processes
on the longitudinal development of air showers,
notably the atmospheric depth of the shower maximum $X_\text{max}$.
Specifically, the two Lorentz-violating decay processes considered
are photon decay
into an electron-positron pair and modified neutral-pion decay
into two photons.
We use Monte Carlo simulations performed with the
CONEX code which was extended to include
these two Lorentz-violating decay processes
at a magnitude allowed by the best previous Earth-based bound.
Compared to standard physics, these Lorentz-violating decay processes
reduce the average $X_\text{max}$
for showers with primary energies above $10^{18}\;\text{eV}$
by an amount that is significantly larger
than the average resolution of current air shower experiments.
Comparing the simulations of the average $X_\text{max}$
to observations, new Earth-based bounds on this
type of Lorentz violation are obtained, which are better
than the previous bounds by more than three orders of magnitude.
Prospects of further studies are also discussed.
\end{abstract}

%CONCEPTS:
%Research Areas: Extensions of gauge sector
%Properties: Lorentz symmetry (Primary)

\maketitle

\section{Introduction}

Ever since its inception, the
standard model (SM) of elementary particle physics
has been extremely successful with its predictions
tested to high precision. However, it is well known that the SM
is not complete, as it does not describe gravity or
dark matter, for example. Current approaches to
establish a comprehensive and fundamental theory
allow for deviations from exact Lorentz symmetry. The
determination of some of the best current bounds on Lorentz violation
in the various sectors of the SM has taken advantage of the
high energies of cosmic rays and gamma rays
(see, e.g., Refs.~\cite{KlinkhamerRisse2008a, KlinkhamerRisse2008b,KlinkhamerSchreck2008} for three research papers
and Refs.~\cite{JacobsonLiberatiMattingly2006,KosteleckyRussell2011}
for two reviews).

We study the impact of Lorentz violation (LV)
on extensive air showers initiated
by cosmic rays in the Earth's atmosphere with a focus on
ultrahigh energies  above $\unit[1]{EeV} = \unit[10^{18}]{eV}$.
In particular, we consider isotropic
nonbirefringent LV in the photon sector, specializing to the case of a photon velocity larger than the maximum attainable velocity of the standard fermions. This approach was explored
in Ref.~\cite{DiazKlinkhamerRisse2016}, where an analytical
\textit{Ansatz}  was used which
modifies the well-known Heitler model for electromagnetic cascades
by including Lorentz-violating photon decays. A significant impact on the longitudinal shower development of electromagnetic cascades was found.

Here, we build upon that previous
work and extend it in several essential ways.
We employ a full Monte Carlo (MC) procedure, which allows us to
study in detail the impact of LV not only on purely electromagnetic
cascades but also on air showers initiated by hadrons.
In addition to the Lorentz-violating decays of
secondary photons with energies above the threshold,
also a modification of the decays of neutral pions
has to be accounted for in the case of hadron primaries.
With these extensions, we are then able to compare the
theoretical expectations to shower observations.

The theory background on LV
in the context of our study is briefly summarized in
Sec.~\ref{sec:theory}. The results of our MC study are presented in
Sec.~\ref{sec:analysis}, in particular the changes in the average
atmospheric depth of the shower maximum $\left<X_\text{max}\right>$ due to LV for purely electromagnetic cascades in Sec.~\ref{subsec:photons}
(in order to compare with Ref.~\cite{DiazKlinkhamerRisse2016})
and for hadron-induced air showers in Sec.~\ref{subsec:hadrons}.
Comparing the latter simulations to $\left<X_\text{max}\right>$ observations, we present an improved bound on this type of LV (Sec.~\ref{subsec:data}).
While the focus of the present work is on $\left<X_\text{max}\right>$,
the impact of the LV modifications on other shower observables
(shower fluctuations and muon content) is briefly
discussed in Sec.~\ref{subsec:others}.
Section~\ref{sec:outlook} contains a summary and an outlook
on future prospects.
Appendix~\ref{app:Lorentz-violating-photon-decays}
gives some details of the Lorentz-violating photon decays
possibly occurring in the extensive air showers.

%%\newpage%%tmp
\section{Theory}
\label{sec:theory}

In a relatively simple extension of standard quantum electrodynamics
(QED), a single term which breaks
Lorentz invariance but preserves CPT and gauge invariance is added to
the Lagrange density~\cite{ChadhaNielsen1983,KosteleckyMewes2002},
\begin{eqnarray}\label{eq:lagrange-dens}
\mathcal{L}(x)&=&
-\frac{1}{4}F^{\mu\nu}(x)F_{\mu\nu}(x) +
\overline{\psi}(x)\left(\gamma^\mu\big[i\partial_\mu-eA_\mu(x)\big]
-m\right)\psi(x)
\nonumber\\[1mm]&&
-\frac{1}{4}(k_F)_{\mu\nu\rho\sigma}F^{\mu\nu}(x)F^{\rho\sigma}(x)\,,
\end{eqnarray}
where the first two terms on the right-hand side correspond to standard
QED and the last term gives CPT-invariant Lorentz violation
in the photon sector
[the CPT transformation corresponds to the
combined operation of charge conjugation (C), parity reflection (P),
and time reversal (T)].
The Maxwell field strength tensor is defined as usual,
$F_{\mu\nu} \equiv \partial_\mu A_\nu - \partial_\nu A_\mu$.
Throughout this article, we use the Minkowski metric
$\eta_{\mu\nu} =[\text{diag}(+1,-1,-1,-1)]_{\mu\nu}$
and its inverse $\eta^{\mu\nu}$ to lower and raise spacetime indices.
In addition, we employ natural units with $\hbar =  c = 1$.

The fixed constant ``tensor'' $(k_F)_{\mu\nu\rho\sigma}$
in \eqref{eq:lagrange-dens}  has 20
independent components, ten of which produce birefringence and eight
of which lead to direction-dependent modifications of the photon
propagation. The remaining two components correspond to an isotropic
modification of the photon propagation and an unobservable double
trace that changes the normalization of the photon field. Isotropic
nonbirefringent LV in the photon sector is then controlled by a single
dimensionless parameter $\kappa$, which enters the fixed tensor $k_F$
from \eqref{eq:lagrange-dens} in the following way:
\begin{subequations}\label{eq:kF-Ansatz}
\begin{eqnarray}
\label{eq:nonbirefringent-ansatz}
(k_F)_{\mu\nu\rho\sigma} &=&
\frac{1}{2} \big(\,
\eta_{\mu\rho}\,\widetilde{\kappa}_{\nu\sigma} -
\eta_{\mu\sigma}\,\widetilde{\kappa}_{\nu\rho} +
\eta_{\nu\sigma}\,\widetilde{\kappa}_{\mu\rho} -
\eta_{\nu\rho}\,\widetilde{\kappa}_{\mu\sigma}  \,\big) \,,
\\[2mm]
\label{eq:isotropic-ansatz}
\widetilde{\kappa}_{\mu\nu}
&=& \frac{\kappa}{2}\left[\text{diag}(3,1,1,1)\right]_{\mu\nu}\,,
\end{eqnarray}
\end{subequations}
where \textit{Ansatz} \eqref{eq:nonbirefringent-ansatz}
gives nonbirefringence
and \textit{Ansatz} \eqref{eq:isotropic-ansatz} isotropy.
From microcausality and unitarity, there is the following
restriction on the ``deformation parameter'' $\kappa$ of the photon
theory~\cite{KlinkhamerSchreck2011}:
$\kappa$ can only take values in the half-open interval $(-1,\,1]$.
Note that the parameter $\kappa$ of the
\textit{Ansatz} \eqref{eq:kF-Ansatz} is denoted by
$\widetilde{\kappa}_\text{tr}$ in, e.g.,
Refs.~\cite{KlinkhamerSchreck2008,KosteleckyRussell2011,KosteleckyMewes2002,%
KlinkhamerSchreck2011}.

The photon propagation is determined by the field equations
obtained from \eqref{eq:lagrange-dens} and \eqref{eq:kF-Ansatz}.
The phase velocity of the photon is found to be given by
\begin{equation}
v_\text{ph} = \frac{\omega}{|\vec{k}|} =
\sqrt{\frac{1-\kappa}{1+\kappa}}\, c\,,
\label{eq:phasevelocity}
\end{equation}
where the velocity $c$ corresponds to the maximum attainable velocity
of the massive Dirac fermion of \eqref{eq:lagrange-dens}.
The photon velocity \eqref{eq:phasevelocity}
is smaller (larger) than $c$ for positive (negative) values of
$\kappa$.
Indeed, the ``operational definition'' of the
LV parameter $\kappa$ is the relative difference of
the squared maximum attainable fermion velocity
($c$, now written as $v_\text{fermion,\,max}$)
and the squared photon velocity ($v_\text{ph}$, now written as $v_\text{photon}$),
\begin{equation}
\kappa
 = \frac{(v_\text{fermion,\,max})^2 - (v_\text{photon})^2}
        {(v_\text{fermion,\,max})^2 + (v_\text{photon})^2}
 \sim 1 -  \frac{v_\text{photon}}{v_\text{fermion,\,max}}
 \,.
\label{eq:kappa-operational-definition}
\end{equation}
As mentioned above,
we set $v_\text{fermion,\,max}\equiv c=1$ by using natural units.

For nonzero values of $\kappa$, certain decay processes that are
forbidden in the conventional Lorentz-invariant theory become
allowed.
Theoretically, these nonstandard decays are rather
subtle; see Ref.~\cite{KaufholdKlinkhamer2005}
for a general discussion
and Refs.~\cite{KlinkhamerSchreck2008,DiazKlinkhamer2015,Klinkhamer2016}
for detailed calculations of the two decay processes considered in this article.

In fact, we now focus on
theory \eqref{eq:lagrange-dens}--\eqref{eq:kF-Ansatz}
for the case of negative $\kappa$, that is, having
a ``fast'' photon, compared to the maximum
attainable velocity of the standard Dirac fermion.
For sufficiently high energy, this photon can then decay
into an electron-positron pair,
\begin{equation}
\label{eq:photon-decay}
\widetilde{\gamma}\to e^{-} + e^{+}\,,
\end{equation}
where $\widetilde{\gamma}$ denotes the nonstandard photon
as described by the Lagrange density \eqref{eq:lagrange-dens}
with \textit{Ansatz} \eqref{eq:kF-Ansatz}.
Specifically, the energy threshold for photon decay
is given by~\cite{KlinkhamerSchreck2008}
\begin{equation}
E^\text{thresh}_\gamma(\kappa) = 2\,m_e\,\sqrt{\frac{1-\kappa}{-2\kappa}}
\sim
\frac{2\,m_e}{\sqrt{-2\kappa}}\,,
\label{eq:photonthreshold}
\end{equation}
in terms of the electron rest mass $m_e \approx \unit[0.511]{MeV}$.
At tree level, the exact decay rate of photon
decay (PhD) as a function of the photon energy $E_\gamma \geq
E^\text{thresh}_\gamma$ has been
calculated~\cite{KlinkhamerSchreck2008,DiazKlinkhamer2015},
\begin{equation}
\Gamma_\text{PhD}(E_\gamma) = \frac{\alpha}{3}\,
\frac{-\kappa}{1-\kappa^2}\,
\sqrt{(E_\gamma)^2-(E^\text{thresh}_\gamma)^2}\;
\Big(2+(E^\text{thresh}_\gamma/E_\gamma)^2\Big)\,,
\label{eq:photondecayrate}
\end{equation}
with the fine-structure constant $\alpha \equiv e^2/(4\pi) \approx 1/137$
and $E^\text{thresh}_\gamma$ from \eqref{eq:photonthreshold}.

The photon decay length drops to scales of a centimeter
right above threshold (cf. Fig.~7 of Ref.~\cite{DiazKlinkhamer2015})
and the decay process resembles a quasi-instantaneous conversion
of photons into electron-positron pairs.
Therefore, ground-based Cherenkov-telescope observations
of gamma rays with energies of order $\unit[10]-\unit[100]{TeV}$
can be used to impose bounds on negative $\kappa$,
with the most stringent Earth-based bound up to now~\cite{KlinkhamerSchreck2008}
\begin{equation}
\label{eq:previousbound}
\kappa > -9 \times 10^{-16} ~~~  \text{($\unit[98]{\%}$ C.L.)}\,.
\end{equation}
It may be of interest to mention that there is also
a qualitative astrophysics bound~\cite{Altschul2005}
at the level of $-10^{-19}$, but this bound is less reliable
than \eqref{eq:previousbound} for reasons explained
in Ref.~\cite{DiazKlinkhamerRisse2016}.

In order to
improve upon bound \eqref{eq:previousbound}, photons of higher energy
than currently observed would be needed. So far, despite extensive
searches for astrophysical (primary) photons with PeV or EeV energies, no unambiguous photon
detection could be reported at these energies
(see, e.g., Ref.~\cite{Niechciol2017}).
However, photons with energies far above $\unit[100]{TeV}$
are expected to be produced as \emph{secondary} particles
when an ultrahigh-energy hadron enters the Earth's atmosphere and initiates an
air shower. In the first interaction of the primary hadron with an
atmospheric nucleus, mostly charged and neutral pions are
produced. The charged pions further interact with particles from the
atmosphere, producing more pions, while the neutral pions, in standard
physics, rapidly
decay into pairs of photons, which in turn trigger electromagnetic
sub-showers. Especially in the start-up phase of the air shower,
where the energies of the secondary particles are still very high,
a modification of the particle dynamics
due to LV can drastically change the overall development of the air shower, as shown in Ref.~\cite{DiazKlinkhamerRisse2016}
for electromagnetic cascades with the immediate decay of above-threshold photons.

Since we intend to study air showers induced by hadrons, we have to
take into account possible related modifications of other processes due to LV,
which may also have an influence on the development of the air
shower. The relevant process here is the decay of the neutral pion
into two nonstandard photons~\cite{DiazKlinkhamerRisse2016,Klinkhamer2016},
\begin{equation}
\label{eq:pi0-decay}
\pi^{0}\to\widetilde{\gamma}+\widetilde{\gamma}\,.
\end{equation}
In the context of the theory considered in this work,
the decay time $\tau$ of the neutral pion is modified by a factor depending on the energy $E_{\pi^{0}}$ of the pion and
the negative LV parameter $\kappa$:
\begin{equation}
\tau(E_{\pi^{0}},\kappa) = \frac{\tau_\text{SM}}{g(E_{\pi^{0}},\kappa)}\,,
\label{eq:piondecaytime}
\end{equation}
with $\tau_\text{SM}$ denoting the neutral-pion decay time in the conventional Lorentz-invariant theory (standard model)
and $g(E_{\pi^{0}},\kappa)$ given by~\cite{Klinkhamer2016}
\begin{equation}
g(E_{\pi^{0}},\kappa) =  \begin{cases}
\frac{\textstyle{\sqrt{1-\kappa^2}}}{\textstyle{(1-\kappa)^3}}
    \left[1-
    \frac
    {\textstyle{\left(E_{\pi^{0}}\right)^2-\left(m_{\pi^{0}}\right)^2}}
    {\textstyle{\left(E^\text{cut}_{\pi^{0}}\right)^2-\left(m_{\pi^{0}}\right)^2}}
    \right]^2,
    & \text{for } \textstyle{E_{\pi^{0}}<E^\text{cut}_{\pi^{0}}},\\
    0\,, & \text{otherwise}\,.
  \end{cases}
\label{eq:gfactor}
\end{equation}
The pion cutoff energy $E^\text{cut}_{\pi^{0}}$ in \eqref{eq:gfactor} is
\begin{equation}
E^\text{cut}_{\pi^{0}} =
m_{\pi^{0}}\, \sqrt{\frac{1-\kappa}{-2\kappa}}
\sim \frac{m_{\pi^{0}}}{\sqrt{-2\kappa}}
\sim \frac{m_{\pi^{0}}}{2m_e} E^\text{thresh}_\gamma
\approx 132\, E^\text{thresh}_\gamma\,,
\label{eq:pionthreshold}
\end{equation}
where the asymptotic
photon-threshold expression \eqref{eq:photonthreshold}
was used
and the numerical values of the neutral-pion and electron rest masses,
$m_{\pi^{0}} \approx \unit[135]{MeV}$
and $m_e \approx \unit[0.511]{MeV}$.
Thus, neutral pions start to be affected significantly at energies
about two orders of magnitude above the photon-decay threshold $E^\text{thresh}_\gamma$.

%%\newpage%%tmp
\section{Analysis}
\label{sec:analysis}

In the following, we discuss the results of our LV study using a full MC
approach. Since we are mainly interested in changes of the
longitudinal development of air showers, we use the MC code CONEX v2r5p40~\cite{BergmannEngelHeckKalmykovOstapchenkoPierogThouwWerner2007,PierogAlekseevaBergmannChernatkinEngelHeckKalmykovMoyonOstapchenkoThouwWerner2006}, which we extend to
include the decay of photons as well as the modified decay of
neutral pions. Photon decays into electron-positron pairs
are implemented as the immediate decay of
photons above the threshold given by \eqref{eq:photonthreshold},
with the energy distributions of the secondary particles modeled
according to Ref.~\cite{DiazKlinkhamer2015}.
An example of the energy distribution is given in
Fig.~\ref{fig:energydistribution} of App.~\ref{app:Lorentz-violating-photon-decays}.
For the modified decay time of neutral pions,
the results \eqref{eq:piondecaytime}--\eqref{eq:pionthreshold}
have been incorporated into the CONEX code.

To describe hadronic interactions in the MC simulation of the air
showers, we use as default the EPOS LHC model~\cite{PierogKarpenkoKatzyYatsenkoWerner2015}
(see below for discussion of other models).
For all other settings, we use the defaults provided by the CONEX code.

%%\newpage%%tmp
\subsection{Primary photons}
\label{subsec:photons}

In order to compare the results of the full MC simulation to those
of the analytical Heitler-type model of Ref.~\cite{DiazKlinkhamerRisse2016}, we first consider
the case of primary photons.
In Fig.~\ref{fig:photons}, the average atmospheric depth of the shower
maximum $\left<X_\text{max}\right>$ is shown as a function
of the primary energy for different values of $\kappa$,
including $\kappa = -9 \times 10^{-16}$ corresponding to
the maximum LV allowed so far
by the previous bound \eqref{eq:previousbound}.
Overall, the MC results confirm
the significant reduction of $\left<X_\text{max}\right>$ and the change of elongation rate
(increase of $\left<X_\text{max}\right>$ with energy)
as expected from Ref.~\cite{DiazKlinkhamerRisse2016}.
For instance, $\left<X_\text{max}\right>$ at $\unit[10^{19}]{eV}$
is reduced by some $\unit[320]{g\,cm^{-2}}$.
The full MC simulation shows a larger decrease compared to
the analytical Heitler-type model.
This can be understood as follows:
the previous analytical approach neglected the fact
that showers initiated
by electrons/positrons are shorter than those initiated by photons of
the same energy.
A further difference comes from using
realistic energy distributions, compared
to the equal-energy-sharing assumption employed
in the analytical estimate
of Ref.~\cite{DiazKlinkhamerRisse2016}.

The substantial reduction in $\left<X_\text{max}\right>$
of approximately $\unit[55]{g\,cm^{-2}}$ just above the threshold energy
for photon decay into an electron-positron pair is due to two effects:
first, having an electron/positron as primary instead of a photon,
which leads to a shower shortened by approximately $\unit[30]{g\,cm^{-2}}$,
and,
second, having two lower-energy primaries instead of one higher-energy
primary, which again leads to a shorter shower.
Since a symmetric share of energies is favored just above threshold
(see Fig.~\ref{fig:energydistribution} of
App.~\ref{app:Lorentz-violating-photon-decays}), the latter effect can be estimated as approximately
(85\,$\times \log_{10} 2)\,$g\,cm$^{-2} \approx \unit[25]{g\,cm^{-2}}$
(using an elongation rate of approximately
$\unit[85]{g\,cm^{-2}}$ per decade for electromagnetic showers).

For energies just above threshold, there is a somewhat
larger elongation rate than at higher energies.
The reason is that, up to about twice the threshold energy,
the primary energy typically allows for just a single case of photon decay.
Only at larger energies, can photon decay happen more than once, leading to a reduction of the elongation rate.

We can also try to obtain a better understanding of
the magnitude of the $\left<X_\text{max}\right>$
reduction at primary energies far above the photon-decay threshold.
In the cascade, a large number of secondary shower photons
will then be produced above threshold and undergo photon decay.
A simplified description is that each conversion
of a secondary photon into an electron-positron pair will reduce the corresponding
sub-shower in a similar manner as observed for the primary conversion.
A difference is due to the energy distribution of the electron-positron pair
which favors a more asymmetric share if the photon energy surpasses
the threshold energy by a factor of 2 or more. Correspondingly,
the effect of a shorter shower due to the production of lower-energy
particles is reduced to a value of approximately $\unit[15]{g\,cm^{-2}}$
(instead of approximately $\unit[25]{g\,cm^{-2}}$ just above threshold)
and leads to a net reduction of approximately
(30+15)\,g\,cm$^{-2} = \unit[45]{g\,cm^{-2}}$ for a sub-shower.
The relative contribution of this sub-shower to the
total shower may be approximated by the fractional energy $f_i$ carried
by the secondary photon relative to the primary energy. The total
reduction of $\left<X_\text{max}\right>$ of the whole shower
is then expected to contain the sum over all sub-showers from photon decay,
\begin{equation}
\Delta \left<X_\text{max}\right>^\text{expected}
  \approx
  \unit[55]{g\,cm^{-2}}
  +  \unit[45]{g\,cm^{-2}}\, \left( \sum f_i - 1 \right)\,,
\label{eq:deltaxmaxexpected}
\end{equation}
where the first term on the right-hand side
comes from the primary conversion and the second
term accounts for secondary conversions.
An explicit study of this argument has been performed
for a primary photon energy of $10^{18}$~eV.
An example of a typical $f_i$ distribution in a shower is shown in
Fig.~\ref{fig:energyfraction} of App.~\ref{app:Lorentz-violating-photon-decays}.
For this case, $\sum f_i \approx 5.5$ gives
$\Delta \left<X_\text{max}\right>^\text{expected}
\approx \unit[258]{g\,cm^{-2}}$,
in reasonable agreement with the result of the MC simulation
in Fig.~\ref{fig:photons}.

%%\newpage%%tmp
\subsection{Primary hadrons}
\label{subsec:hadrons}

We now consider the case of secondary photons produced in air showers
initiated by primary hadrons. This has not been studied before and requires an account of the modification
of the neutral-pion decay as well (see Sec.~\ref{sec:theory}).
Compared to the case of primary photons, where already the initial particle
is modified by LV, a smaller impact on $\left<X_\text{max}\right>$
is expected here, as hadronic interactions (dominating the start of the
cascade) are unaffected by LV in the photon sector.

\begin{figure}[p]  %[t]
\centering
\includegraphics[width=0.75\textwidth]{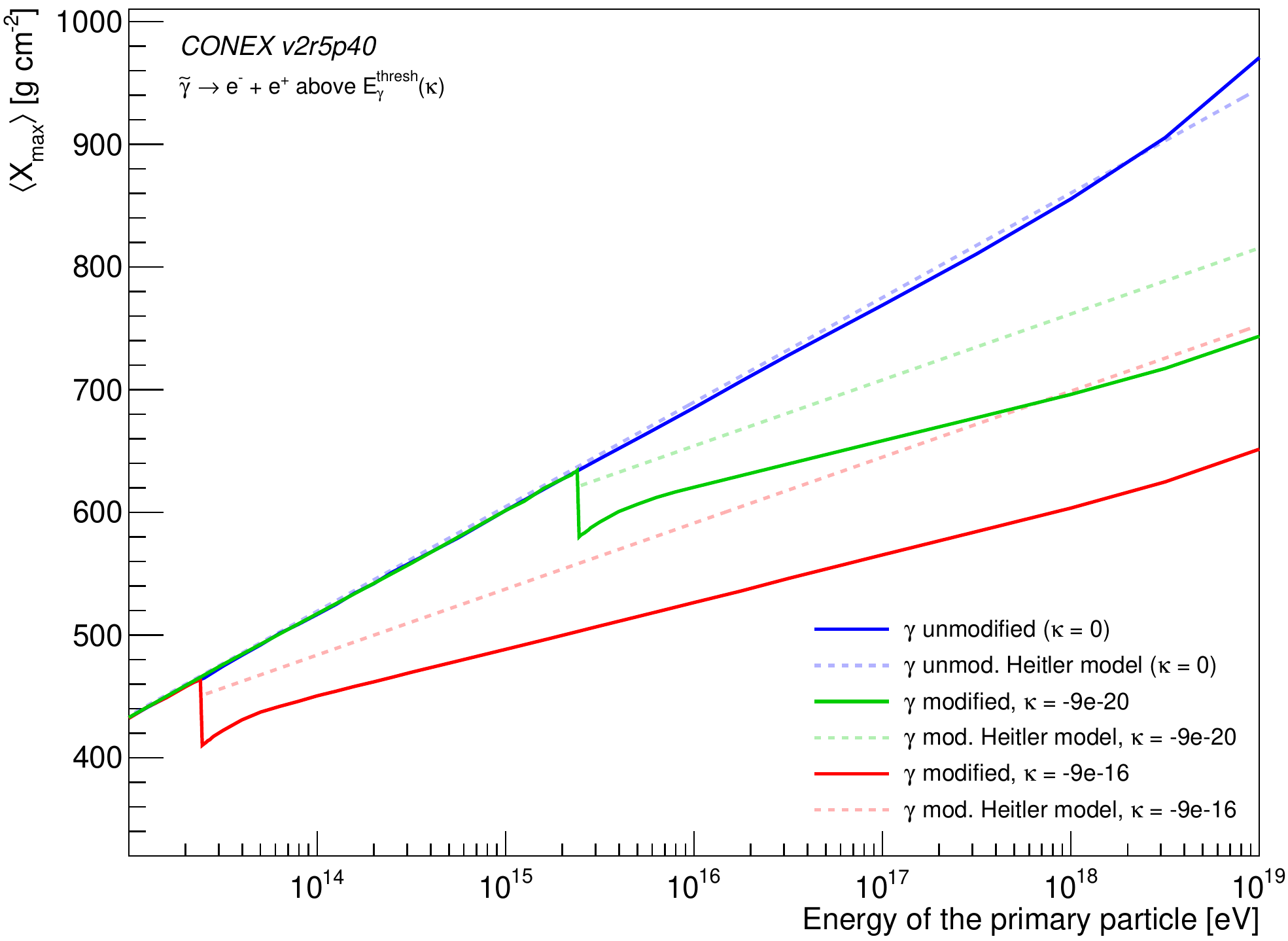}
%{lvstudy_photon_v095.pdf}
\caption{Average atmospheric depth of the shower
maximum $\left<X_\text{max}\right>$ as a function of the primary energy of
a primary photon, taken from MC simulations performed with the
CONEX code which was modified to include Lorentz violation
controlled by a negative parameter $\kappa$.
The dashed lines indicate the
$\left<X_\text{max}\right>$ values
expected from the analytical Heitler-type
model of Ref.~\cite{DiazKlinkhamerRisse2016}.}
\label{fig:photons}%%Fig1
%\end{figure}
\vspace*{5mm}
%\begin{figure}[p]  %[t]
\centering
\includegraphics[width=0.745\textwidth]{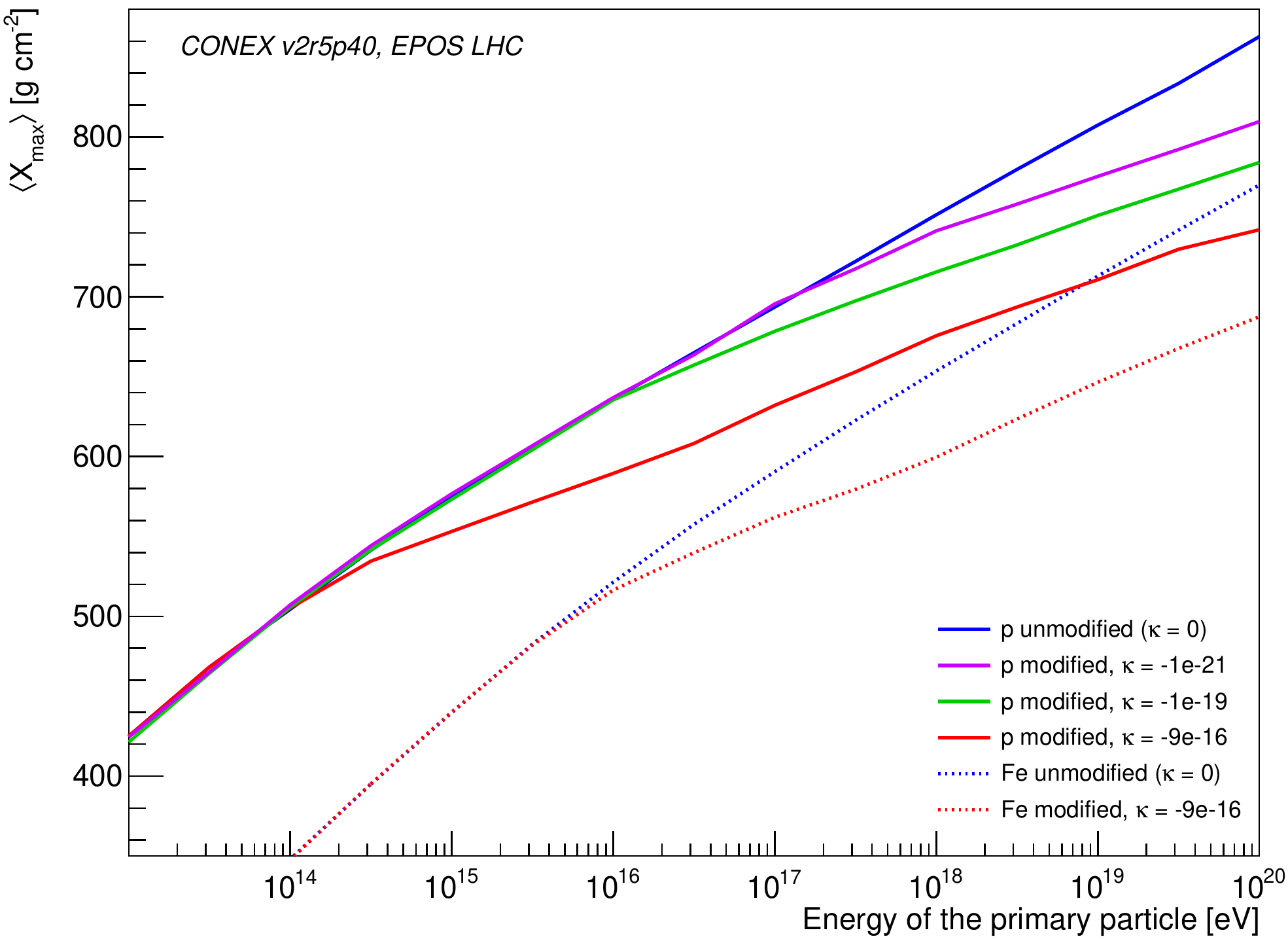}
%{lvstudy_hadrons_xmax_v011.pdf}
\caption{Simulated values of
$\left<X_\text{max}\right>$ as a function of the primary
energy for primary protons and iron nuclei, where
different values of the Lorentz-violating parameter $\kappa$
are considered.}
\label{fig:hadrons}%%Fig2
\end{figure}

The results are displayed in Fig.~\ref{fig:hadrons}
for primary protons and iron nuclei. Also hadron-induced air showers
are seen to be significantly affected in terms of $\left<X_\text{max}\right>$, which is approximately
$\unit[100]{g\,cm^{-2}}$ smaller compared to the unmodified case,
for protons at $\unit[10^{19}]{eV}$ and a $\kappa$
value saturating bound \eqref{eq:previousbound}.
The impact is large with respect to the experimental resolution.
For instance, the $X_\text{max}$ resolution of the Pierre Auger Observatory
is better than $\unit[26]{g\,cm^{-2}}$ at $\unit[10^{17.8}]{eV}$, improving
to about $\unit[15]{g\,cm^{-2}}$ above $\unit[10^{19.3}]{eV}$~\cite{Auger2014}.
Figure~\ref{fig:hadrons} also shows that the
elongation rate is modified as well. The reduction amounts to
about $\unit[25]{g\,cm^{-2}}$ per decade,
relatively independent of  the $\kappa$ value
and the primary type considered.

The modification of $\left<X_\text{max}\right>$ occurs
if the energy per nucleon of the primary particle exceeds
the LV threshold energy $E^\text{thresh}_\gamma$
by a factor of approximately $10$,
because secondary photons above threshold then start to be produced.
Neutral pions become stable at sufficiently high energies
and the change occurs at energies about two orders of magnitude
above the photon-decay threshold energy, according to the
estimate from \eqref{eq:pionthreshold}.
The onset of neutral pions becoming stable is noticed only as a minor
effect on $\left<X_\text{max}\right>$, i.e.,
a slight upturn of the modified curve.
Changes in the values of $\kappa$ and in the type of primary
(e.g., primary iron nuclei instead of protons)
give shifts of the curves in the ways naively expected.

\begin{figure}[p]  %[t]
\centering
\includegraphics[width=0.75\textwidth]{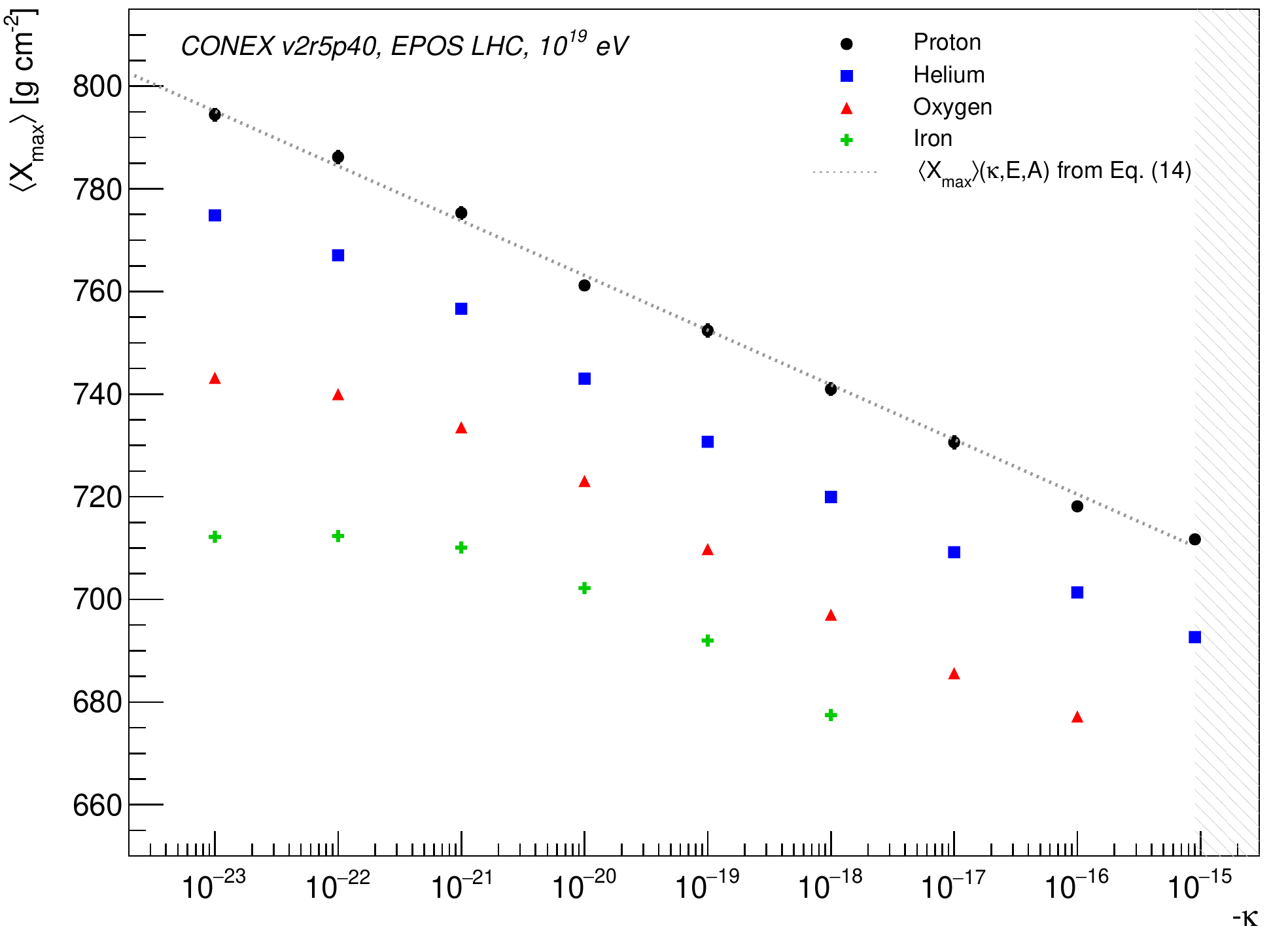}
%{lvstudy_kappascan_v4.pdf}
%{improved-bound-lv-eas-fig3-v1.pdf}
%{lvstudy_kappascan_v092.pdf}
\caption{Simulated values of
$\left<X_\text{max}\right>$ as a function of $-\kappa$ for
different primary nuclei (proton, helium, oxygen, and iron)
at a fixed primary energy of $\unit[10^{19}]{eV}$.
The dotted lines correspond to
the parametrization \eqref{eq:xmaxparametrization} and
the hatched area on the right
indicates the range of $-\kappa$ values that is
excluded by the previous bound \eqref{eq:previousbound}.
}
\label{fig:masses}%%Fig3
%\end{figure}
\vspace*{5mm}
%\begin{figure}[p]  %[t]
\centering
\includegraphics[width=0.75\textwidth]{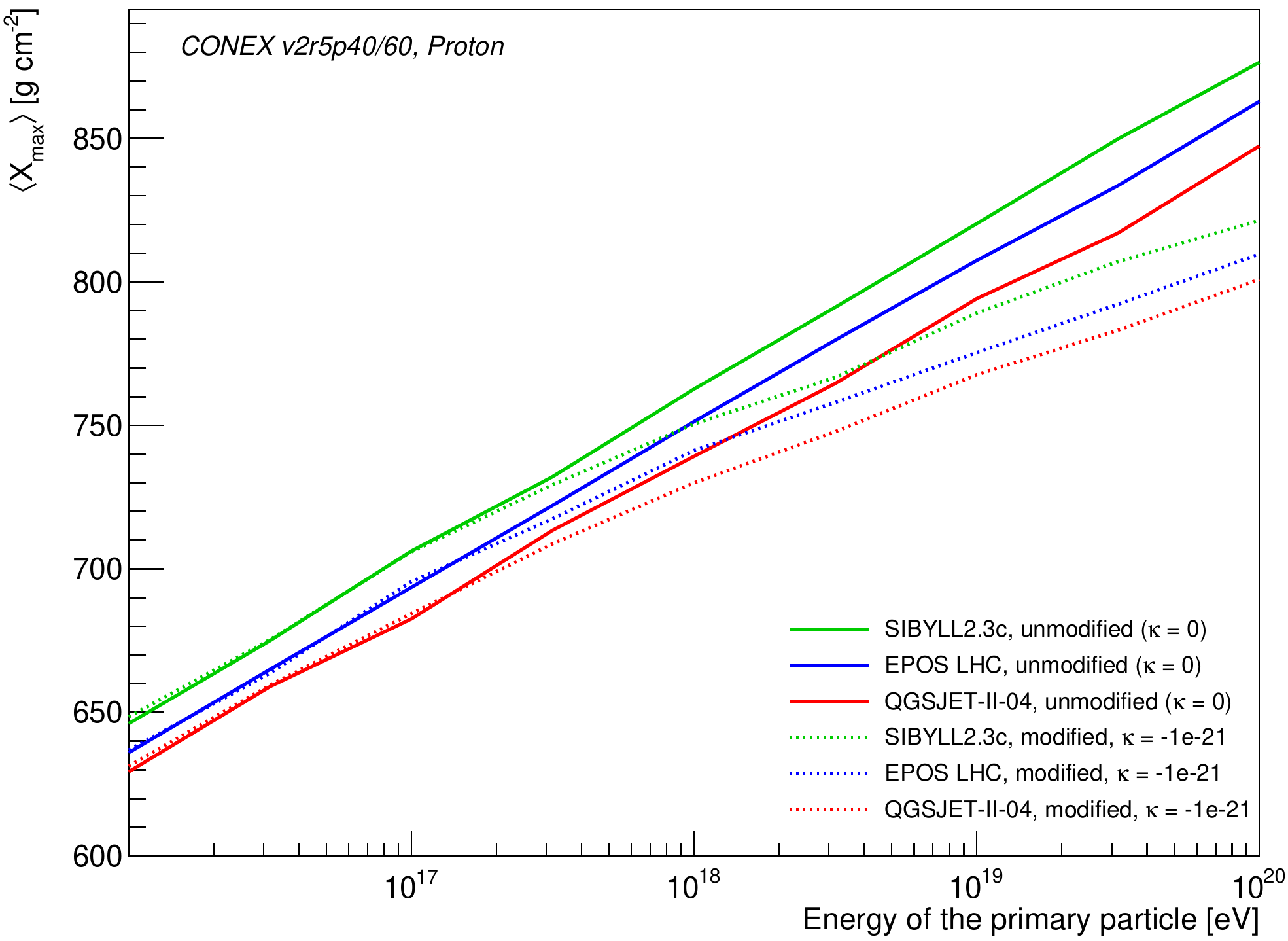}
%{lvstudy_models_v011.pdf}
\caption{Simulated values of
$\left<X_\text{max}\right>$ as a function of the primary
energy for primary protons, where different hadronic-interaction models
are used.}
\label{fig:models}%%Fig4
\end{figure}

Using different primary nuclei (proton, helium, oxygen, and iron), we have determined
the values of $\left<X_\text{max}\right>$ as a function of $\kappa$ at
a fixed primary energy of $\unit[10^{19}]{eV}$.
As shown in Fig.~\ref{fig:masses},
$\left<X_\text{max}\right>$ scales linearly
with $\log_{10}(-\kappa)$ if the photon-decay
threshold energy is well below the energy per nucleon
of the primary particle. This can be used to obtain a
parametrization of $\left<X_\text{max}\right>$ as a function of
the negative LV parameter $\kappa$,
the primary energy $E$, and the primary mass $A$. For EPOS LHC, this
parametrization is given by%
\begin{equation}
\left<X_\text{max}\right>(\kappa, E, A)
\,\Big|^\text{(above-threshold)}
=\,p_0\,+\,p_1\,\log_{10}(-\kappa)\\
+\,p_2\,\Big(\log_{10}(E\,\unit{[eV]})-19\Big)\\
+\,p_3\,\ln(A)\,,
\label{eq:xmaxparametrization}
\end{equation}
where $p_0 = \unit[(550\,\pm\,3)]{g\,cm^{-2}}$ and $p_1 =
\unit[(-10.7\,\pm\,0.1)]{g\,cm^{-2}}$ are determined by a fit to the
proton distribution shown in Fig.~\ref{fig:masses}, $p_2 =
\unit[(34.3\,\pm\,0.3)]{g\,cm^{-2}}$ is the $\kappa$- and $A$-independent elongation rate
taken from Fig.~\ref{fig:hadrons}, and $p_3 = \unit[(-15.8\,\pm\,0.1)]{g\,cm^{-2}}$ is obtained
from a fit to the distributions for the different primary particles
shown in Fig.~\ref{fig:masses}. It should be noted that for the
determination of the improved bound on $\kappa$
(to be discussed in Sec.~\ref{subsec:data}),
only the parametrization of $\left<X_\text{max}\right>$ for the proton
case is needed. For completeness, we have included
the generalization of the parametrization to heavier nuclei.

We have also checked the dependence of the simulations
on the choice of the hadronic-interaction model.
The uncertainty of $\left<X_\text{max}\right>$ due to modeling
hadronic interactions is about $\pm$$\unit[20]{g\,cm^{-2}}$
around the predictions of EPOS LHC~\cite{Pierog2017}. Likewise,
the predictions from the alternative models
QGSJET-II-04~\cite{Ostapchenko2011} and
SIBYLL~2.3c~\cite{AhnEngelGaisserLipariStanev2009,RiehnEngelFedynitchGaisserStanev2015,RiehnDembinskiEngelFedynitchGaisserStanev2015} may be regarded as resembling the lower (QGSJET-II-04) and upper (SIBYLL~2.3c)
range of possible $\left<X_\text{max}\right>$ predictions~\cite{Pierog2017}.
The results for proton primaries using QGSJET-II-04 and SIBYLL~2.3c
are displayed in Fig.~\ref{fig:models}.
As expected, the LV-modified curves reflect the differences between the
unmodified curves for the different hadronic-interaction models.
Differences between the models of the modified $\left<X_\text{max}\right>$
values are smaller compared to the unmodified $\left<X_\text{max}\right>$ values.
At $\unit[10^{19}]{eV}$, for instance, the predictions cover a range of
approximately $\unit[22]{g\,cm^{-2}}$ for the modified  case
($\kappa= -1 \times 10^{-21}$)
and approximately $\unit[28]{g\,cm^{-2}}$
for the unmodified case ($\kappa=0$).
Compared to EPOS LHC, the differences in
$\left<X_\text{max}\right>$ are about $\unit[8-13]{g\,cm^{-2}}$,
where QGSJET-II-04 gives smaller
$\left<X_\text{max}\right>$ values
and SIBYLL 2.3 larger $\left<X_\text{max}\right>$ values.
These differences are small compared
to the overall reduction of $\left<X_\text{max}\right>$ by the LV modification allowed
by previous bound \eqref{eq:previousbound}.

%%\newpage%%tmp
\subsection{Comparison to $\left<X_\text{max}\right>$ measurements}
\label{subsec:data}

\begin{figure}[t]  %[t]
\centering
\vspace*{-0mm}
\includegraphics[width=0.75\textwidth]{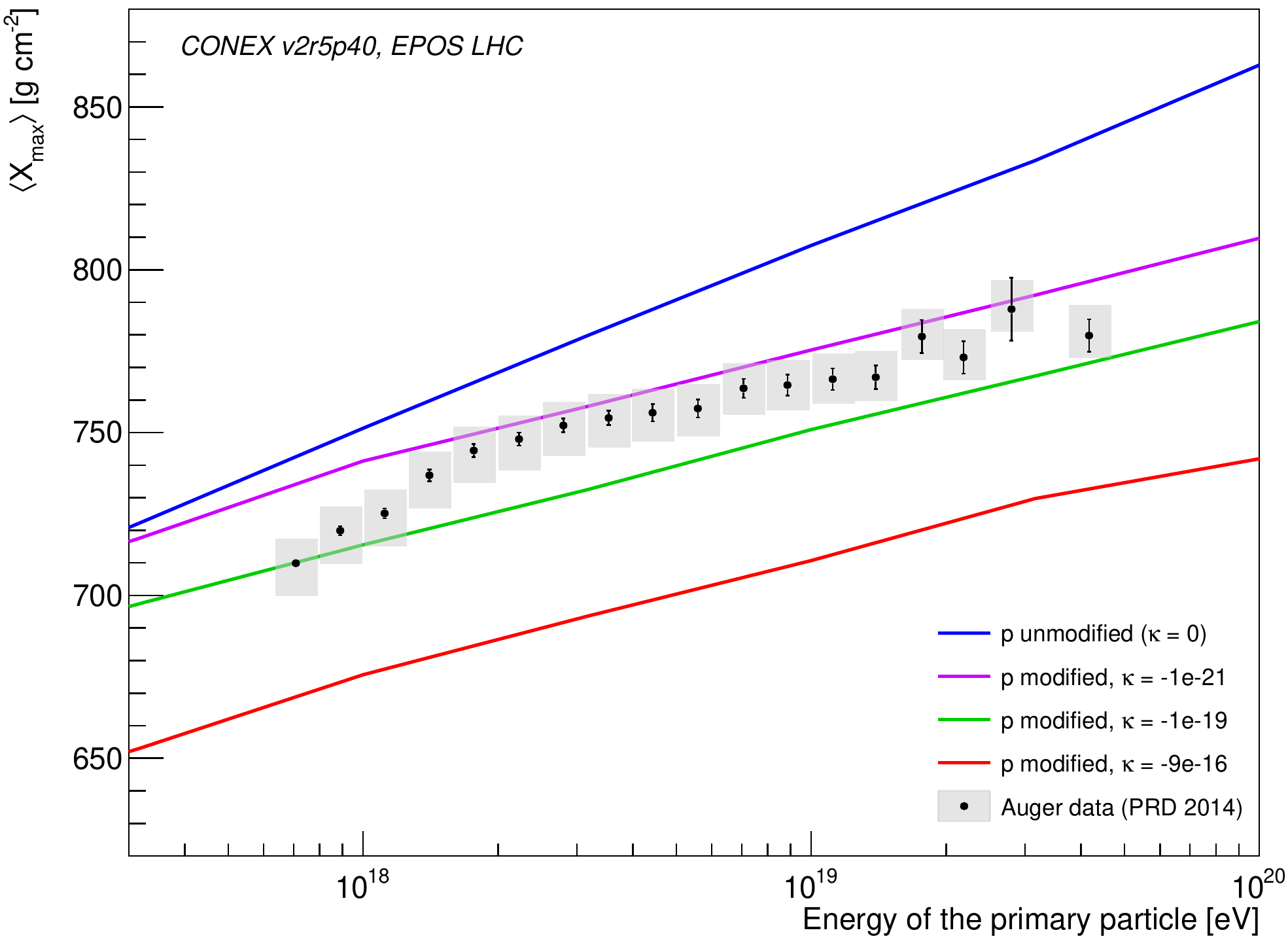}
%{lvstudy_data_v011.pdf}
\vspace*{-0mm}
\caption{Simulated values of
$\left<X_\text{max}\right>$ as a function of the primary
energy for primary protons compared to
measured values of $\left<X_\text{max}\right>$ by
the Pierre Auger Observatory~\cite{Auger2014}. The gray boxes
around the data points indicate the systematic uncertainties of the measurements.}
\label{fig:data}%%Fig5
\end{figure}

A comparison of our $\left<X_\text{max}\right>$ simulations
with $\left<X_\text{max}\right>$ measurements is given in Fig.~\ref{fig:data}.
Only measurements from the Pierre Auger Observatory~\cite{Auger2014}
are shown here, since the $\left<X_\text{max}\right>$ measurements from
the Telescope Array (TA) are not corrected for detector effects.
In any case,
the TA measurements have been found to be consistent with the Auger
measurements~\cite{Unger2015}. For the maximum LV allowed by previous
bound \eqref{eq:previousbound}
from Ref.~\cite{KlinkhamerSchreck2008},
it can be
seen that $\left<X_\text{max}\right>$ predictions are significantly
below the observations, regardless of assumptions on the primary
mass and the interaction model. Thus, stricter constraints than
before can be placed on negative $\kappa$ by the method presented in this article.

Allowing, most conservatively, for the case of a pure proton composition,
one can see that also the $\kappa = -10^{-19}$ case
in Fig.~\ref{fig:data}  appears to be on the lower side of
the observed $\left<X_\text{max}\right>$ values above $\unit[10^{18}]{eV}$.
While we concentrate here on obtaining a bound,
it is interesting to note that the elongation rate
of a constant proton composition
in the modified case for $\kappa \sim -10^{-20}$
turns out to agree reasonably well with the observations
at or above $2\times$10$^{18}$~eV.
Focusing on the energy bin around $\unit[2.8 \times10^{18}]{eV}$
and taking the uncertainties on the measured
$\left<X_\text{max}\right>$ values into account,
we obtain the following bound:
\begin{equation}
\label{eq:newbound}
\kappa > -3 \times 10^{-19} ~~~  \text{($\unit[98]{\%}$ C.L.)}\,.
\end{equation}
For completeness, the corresponding bound at $\text{$\unit[99.9]{\%}$ C.L.}$
is $\kappa > -1.2 \times 10^{-18}$.  %%MR email 14Dec2017 10:50

The bound \eqref{eq:newbound}
is based on EPOS-LHC simulations
and the parametrization~\eqref{eq:xmaxparametrization},
where a systematic uncertainty of $\unit[20]{g\,cm^{-2}}$ has been assumed to account for the
uncertainties related to the description of hadronic interactions.
We note that the value of $\unit[20]{g\,cm^{-2}}$ is conservative,
as it is taken from the
uncertainty of the unmodified simulations, while the uncertainty for the
modified case is likely to be reduced (see Sec.~\ref{subsec:hadrons}).
Without additional systematic model uncertainties,
the $\unit[98]{\%}$ C.L.
bounds derived from SIBYLL~2.3c, EPOS LHC, and QGSJET-II-04
are $-2 \times 10^{-19}$, $-0.2 \times 10^{-19}$,
and $-0.02 \times 10^{-19}$, respectively.
These three bounds are even tighter than
the bound quoted in (\ref{eq:newbound}).

The bound on negative $\kappa$ as given by \eqref{eq:newbound}
improves the previous
bound \eqref{eq:previousbound} by a factor of approximately $3000$.
As a comparison to the previous
approach of Ref.~\cite{KlinkhamerSchreck2008},
the observation of a primary photon with an energy of about
$\unit[2]{PeV}$ $=$ $\unit[2\times 10^{15}]{eV}$
would be needed to get a similar bound on negative $\kappa$.

\begin{figure}[p]  %[t]
\centering
\includegraphics[width=0.745\textwidth]{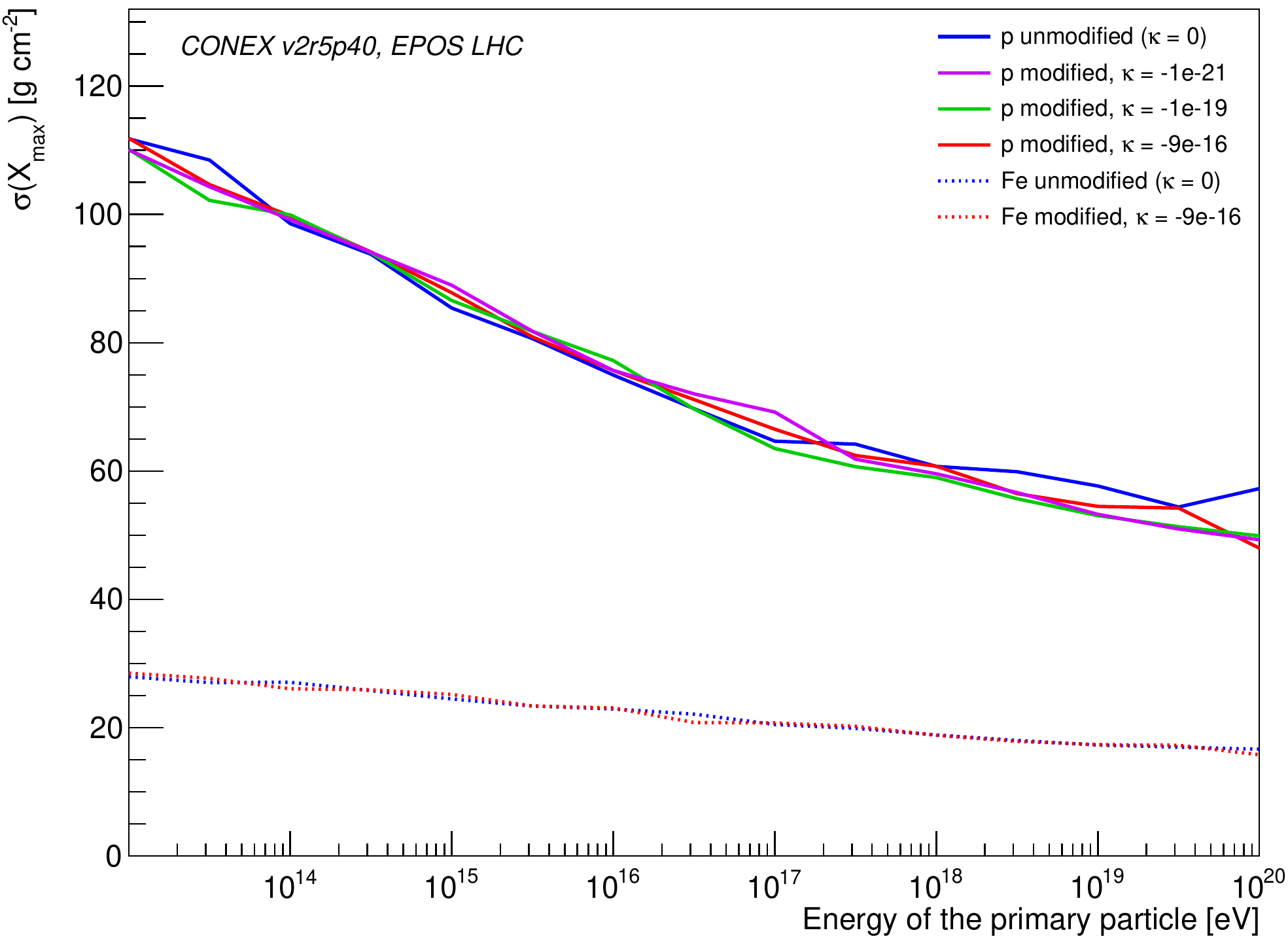}
%{lvstudy_hadrons_rmsxmax_v011.pdf}
\caption{Shower fluctuations $\sigma(X_\text{max})$ as a function of the primary
energy for primary protons and iron nuclei.}
\label{fig:sigma}%%Fig6
%\end{figure}
\vspace*{5mm}
%\begin{figure}[p]
\centering
\includegraphics[width=0.745\textwidth]{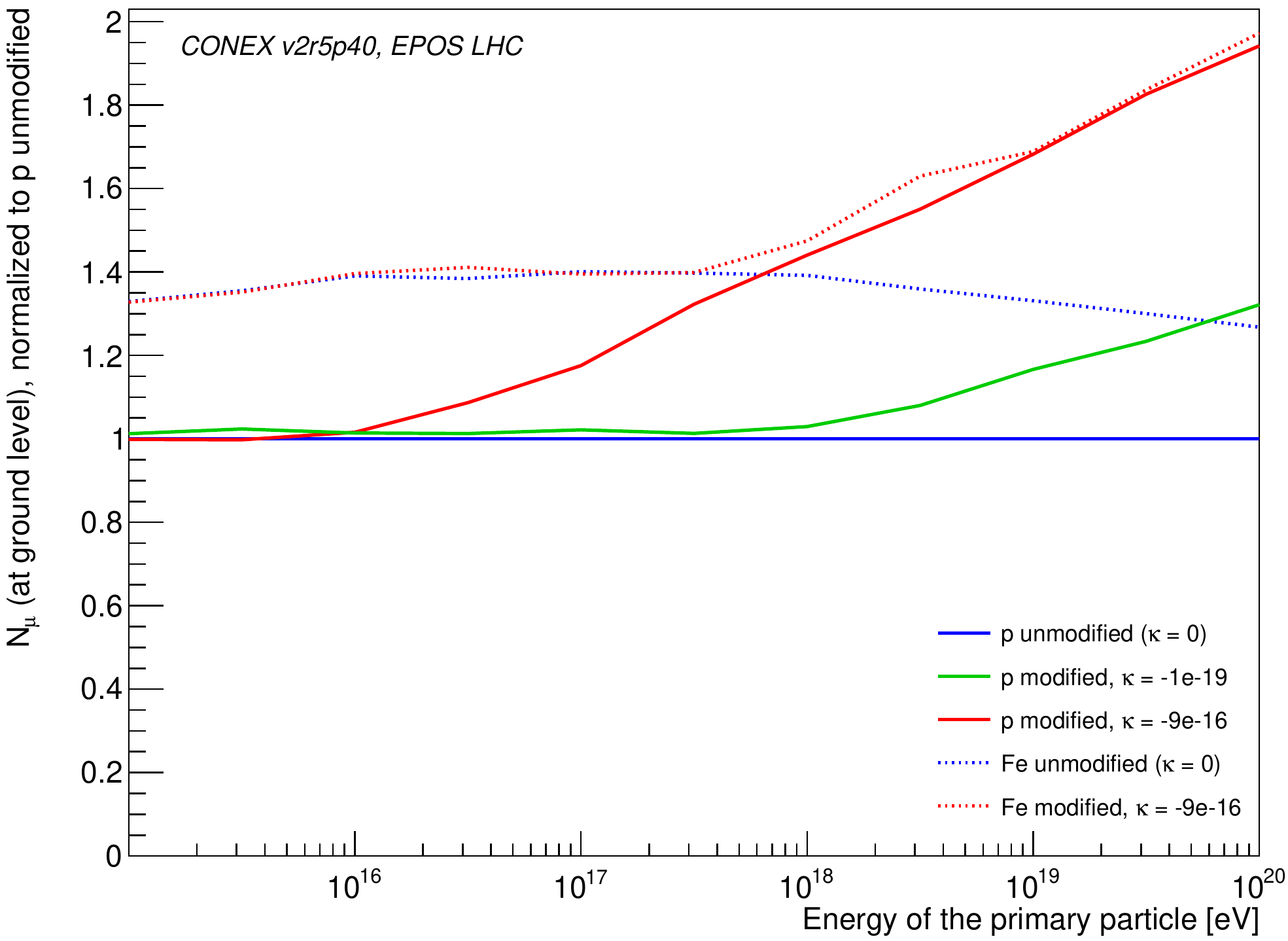}
%{lvstudy_hadrons_muground_normalized_v011.pdf}
\caption{Average number of ground muons, normalized to the case of unmodified
proton primaries, as a function of the primary energy
for primary protons and iron nuclei.}
\label{fig:muons}%%Fig7
\end{figure}

%%\newpage%%tmp
\subsection{Other shower observables}
\label{subsec:others}
\vspace*{-0mm}

While the focus of the present work has been on $\left<X_\text{max}\right>$,
additional shower observables can be studied with the simulation setup.

The impact of the LV modifications on the shower-to-shower fluctuations
$\sigma(X_\text{max})$ is shown in Fig.~\ref{fig:sigma}.
Interestingly, the LV modifications leave $\sigma(X_\text{max})$  essentially unchanged, different from the behavior of $\left<X_\text{max}\right>$.
This is understandable, as the shower fluctuations are dominated by
fluctuations in the first interactions of the highest-energy hadronic primaries, which remain unmodified here
(the LV in the photon sector is expected to lead
to negligible LV loop corrections of the hadronic interactions).

Further, we checked for a possible effect on the muon content of the shower.
Since neutral pions above $E^\text{cut}_{\pi^{0}}$
re-interact,  a corresponding
increase of the number of muons could result
(see also Refs.~\cite{Tomar2017,Boncioli-etal2015}
for considerations of shower muons and LV).
As can be seen in Fig.~\ref{fig:muons},
the number of muons is indeed affected. For primary protons with
$\kappa = -9 \times 10^{-16}$,
the increase of the muon number starts
for primary energies of about a factor
$3$ above the pion cutoff energy
($E^\text{cut}_{\pi^{0}} \approx  \unit[3.2\times10^{15}]{eV}$ in this case).  The muon excess grows with energy, reaching a factor 1.5 in the EeV range.
A similar effect is observed for primary iron nuclei,
with the primary energy
scaled according to the number of nucleons.
Remarkably, the muon number at the highest energies
turns out to be quite similar for proton and
iron primaries in the modified case, in contrast to the unmodified case.
As $\kappa = -9 \times 10^{-16}$ is excluded by
our new bound \eqref{eq:newbound}, Fig.~\ref{fig:muons} also shows the
results for proton primaries at the value $\kappa = -1 \times 10^{-19}$.
The increase of muon number then starts at higher energies.

%%\newpage%%tmp
\vspace*{-5mm}
\section{Summary and outlook}
\label{sec:outlook}
\vspace*{-2mm}

In the present article,
we have considered isotropic
Lorentz violation in a simplified theory of
photons and electrically charged Dirac fermions.
This Lorentz violation is described by a single
dimensionless parameter $\kappa$, whose physical
meaning is clarified by \eqref{eq:kappa-operational-definition}.
Our focus has been on the case of negative $\kappa$ with a ``fast'' photon.

By implementing Lorentz-violating effects
(photon decay and the suppression of neutral-pion decay)
in a full MC shower simulation, we have studied the impact of LV
on air showers initiated by ultrahigh-energy cosmic rays.
This method exploits the expected production of secondary photons with
energies far above $\unit[100]{TeV}$ and the accelerated shower development due to photon decay.
The average depth of the shower maximum $\left<X_\text{max}\right>$
can be reduced by some $\unit[100]{g\,cm^{-2}}$ at $\unit[10^{19}]{eV}$
and the difference increases with
energy due to a reduced elongation rate. This
reduction value of $\unit[100]{g\,cm^{-2}}$ at $\unit[10^{19}]{eV}$
is well above the
typical $X_\text{max}$ resolution of 15$-$$\unit[20]{g\,cm^{-2}}$
in current air shower experiments.
The shower fluctuations $\sigma(X_\text{max})$ are not affected by the
LV modification.
The number of muons at ground level
has been found to increase significantly above the LV cutoff energy of
neutral-pion decay.

With a value of $\kappa = -9 \times 10^{-16}$
as allowed so far by the previous bound~\cite{KlinkhamerSchreck2008},
the predictions of $\left<X_\text{max}\right>$
are at odds with the measurements, irrespective of the primary mass and the interaction model:
much deeper showers
are observed than the showers expected theoretically.
From
$\left<X_\text{max}\right>$ alone, an improved bound of
$\kappa = -3 \times 10^{-19}$ (at $\unit[98]{\%}$ C.L.)
has been obtained in Sec.~\ref{subsec:data}.
For this new bound \eqref{eq:newbound},
the primary composition has, most conservatively,
been assumed
to consist of protons only. Heavier primaries have smaller
$\left<X_\text{max}\right>$ values, which would lead to even stronger bounds.

Remark that the magnitude of the improved
negative-$\kappa$ bound \eqref{eq:newbound}
is only a factor $5$ larger than the
positive-$\kappa$ bound~(15a) from Ref.~\cite{KlinkhamerSchreck2008},
$\kappa < 6 \times 10^{-20} ~  \text{($\unit[98]{\%}$ C.L.)}$.
Future improved negative-$\kappa$ bounds
using only $\left<X_\text{max}\right>$ can come
from further data, also at higher energies, for instance by the present upgrade of the Pierre Auger Observatory~\cite{Engel2015,Auger2015}.
Moreover,
reducing present uncertainties helps: in case of a model uncertainty of
$\unit[15]{g\,cm^{-2}}$, the negative-$\kappa$ bound \eqref{eq:newbound}
improves by a factor of approximately $2$.
A similar effect is obtained by reducing the experimental uncertainty
to half of its present value.

Significantly improved negative-$\kappa$ bounds
appear possible if other observables beyond
$\left<X_\text{max}\right>$ are incorporated.
In fact, a pure proton composition above $\unit[3\times10^{18}]{eV}$,
as conservatively assumed here,
is already excluded by other air shower measurements.
Specifically, observations of $\sigma(X_\text{max})$~\cite{Auger2014}
and observations of the correlation
between $X_\text{max}$ and the ground signal~\cite{Auger2016} show a mixed composition with a
significant fraction of heavier nuclei.
A mixed composition
will provide further improvements of
the negative-$\kappa$ bound \eqref{eq:newbound}.
As illustration, let us assume an
average primary mass  $\left<A\right> \approx 4$.
Then, the resulting predicted
$\left<X_\text{max}\right>$ would coincide with that of primary helium.
For a given observed value of $\left<X_\text{max}\right>$
and according to Fig.~\ref{fig:masses}, this would imply
a further factor of approximately $100$
improvement on the negative-$\kappa$ bound \eqref{eq:newbound}.
We leave this analysis to the future.

\vspace*{-5mm}
\subsection*{Acknowledgments}
\vspace*{-3mm}
We thank T. Pierog for his help in modifying the CONEX
source code.
In addition, we acknowledge useful discussions with
J.S. D\'{i}az. This work was partially supported by the
German Federal Ministry of Education and Research (BMBF),
the Helmholtz Alliance for Astroparticle Physics (HAP),
and the German Research Foundation (DFG).

%%\newpage%%tmp
\vspace*{-5mm}
\begin{appendix}
\section{Lorentz-violating photon decays}
\label{app:Lorentz-violating-photon-decays}
\vspace*{-3mm}

\begin{figure}[h]
\vspace*{-2mm}
\centering
\includegraphics[width=0.75\textwidth]{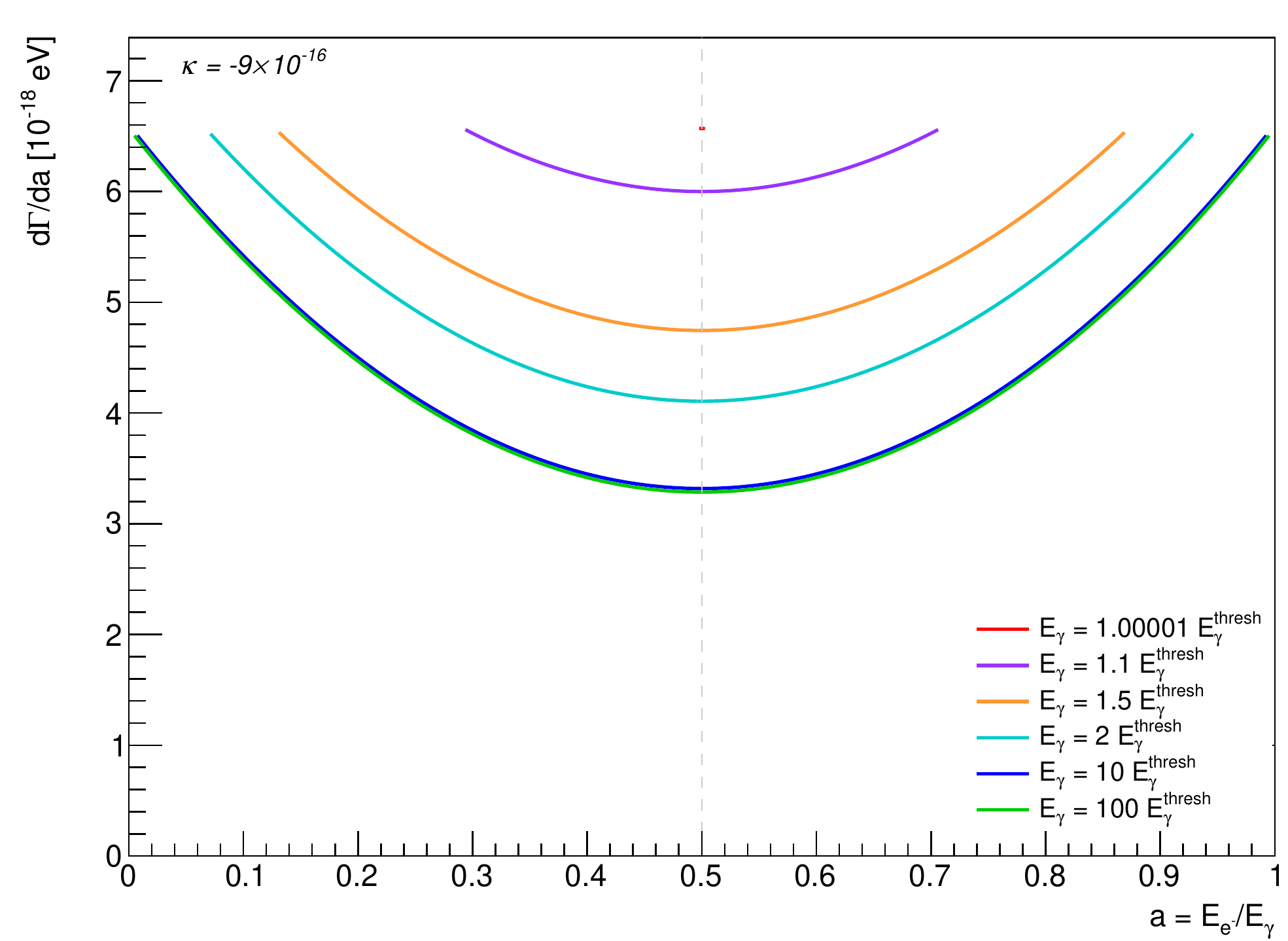}
%{energydistribution_-9e-16_v092.pdf}
\vspace*{-4mm}
\caption{Differential decay rate $\text{d}\Gamma/\text{d}a$
for photon decay into an electron-positron pair as
a function of the fraction $a$ of the energy of the resulting
electron with respect to the initial energy $E_\gamma$ of the
decaying photon. Shown are the decay rates for different
initial energies in units of the threshold energy $E^\text{thresh}_\gamma$
from \eqref{eq:photonthreshold} for the case $\kappa=-9\,\times\,10^{-16}$.}
\label{fig:energydistribution}%%Fig8
\end{figure}

\begin{figure}[h!]
\centering
\includegraphics[width=0.75\textwidth]{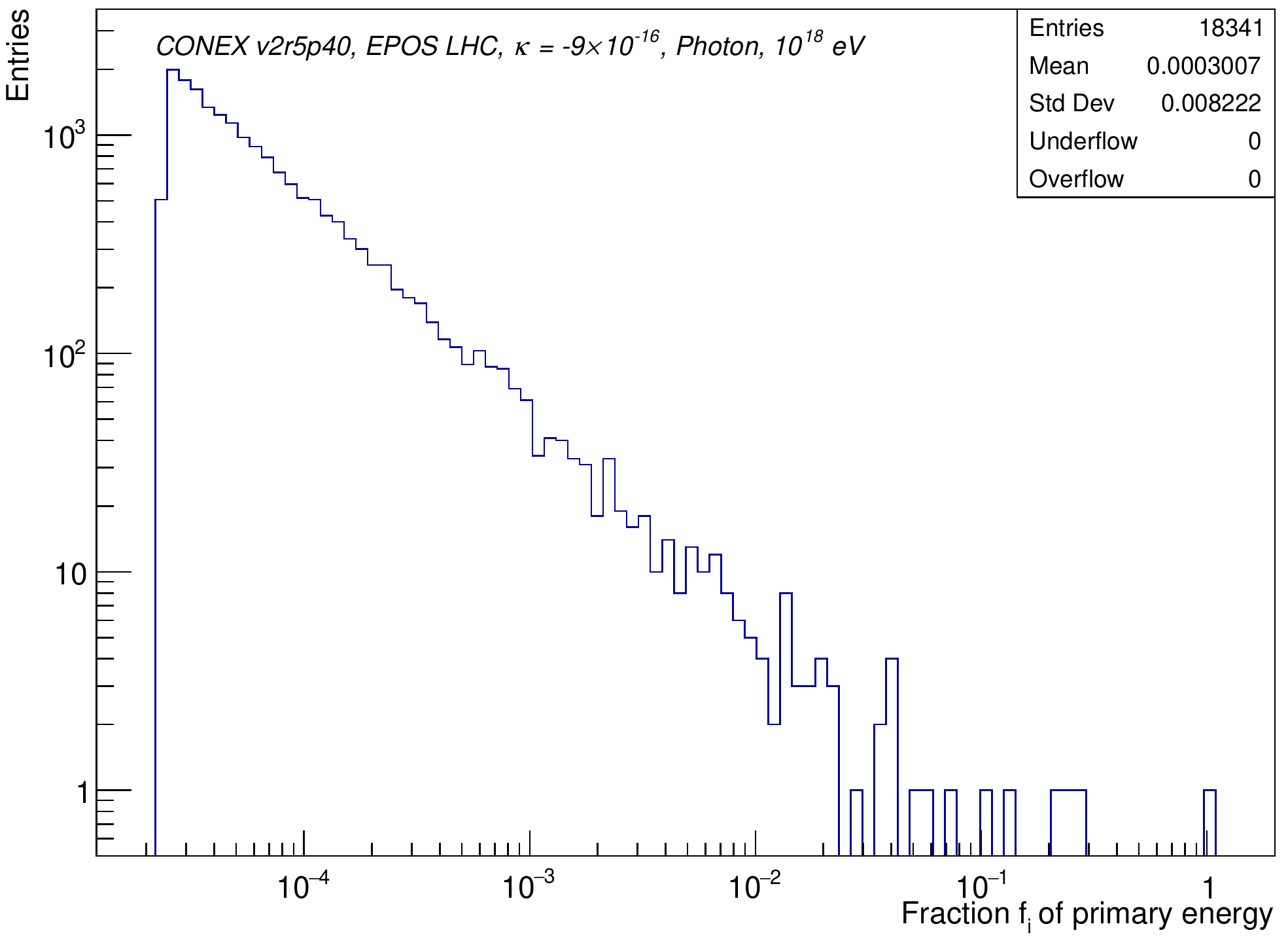}
%{energyfractions_1e18_v095.pdf}
\vspace*{-4mm}
\caption{Distribution of fractions $f_i$ of the energy of decaying photons
relative to the primary energy for a single photon-induced
air shower with a primary energy of $10^{18}\,\text{eV}$.
Each decaying photon in
the shower contributes to the distribution. The decay of the
primary photon is seen as $f_1 = 1$. The minimum value of $f_i$
is determined by the Lorentz-violating
energy threshold \eqref{eq:photonthreshold}.
The simulation has been performed with the modified CONEX code and
Lorentz-violating parameter $\kappa=-9\,\times\,10^{-16}$.}
\label{fig:energyfraction}%%Fig9
\vspace*{0cm}
\end{figure}

In this appendix, we present some further details on the
Lorentz-violating photon decay process, based on the
theoretical results from
Refs.~\cite{KlinkhamerSchreck2008,DiazKlinkhamer2015}.
Figure~\ref{fig:energydistribution} gives the differential decay rate
for photon decay into an electron-positron pair
and Fig.~\ref{fig:energyfraction} shows the energy fraction
of decaying photons.

%\vspace*{4mm}
\end{appendix}

%%\newpage
%\bibliography{improved-bound-lv-eas-v1}
%\bibliography{improved-bound-lv-eas-v2}
%\bibliography{improved-bound-lv-eas-v3}
\bibliography{improved-bound-lv-eas-v4}

\end{document}